\documentclass[3p,10pt,a4paper]{elsarticle}
\usepackage{graphicx}
\usepackage{amssymb}
\usepackage{amsthm}
\usepackage{amsmath}
\usepackage[breaklinks]{hyperref}
\usepackage{xcolor}
\usepackage{lineno}
\modulolinenumbers[1]
\usepackage{balance}
\usepackage{chemfig}
\usepackage{pdfpages}

\journal{Carbon}

\begin{document}

\begin{frontmatter}

%\title[H and H$_2$ collisions with cyclo18carbon]{On the hydrogenation of a cyclo[18]carbon molecule 
%by H and H$_2$ irradiation}
\title{Formation of C$_{18}$H and C$_{18}$H$_2$ molecules by low energy irradiation with atomic and molecular hydrogen}

\author[a]{F. J. Dom\'inguez-Guti\'errez\corref{author}}
\author[b]{C. Mart\'inez-Flores}
\author[c]{P. S. Krstic}
\author[d,e]{R. Cabrera-Trujillo}
\author[a]{U. von Toussaint}

\cortext[author] {Corresponding author: javier.dominguez@ipp.mpg.de}
\address[a]{Max-Planck-Institut f{\"u}r Plasmaphysik, Boltzmannstrasse 2, 85748 Garching, Germany}
\address[b]{Departamento de Qu\'imica, Divisi\'on de Ciencias B\'asicas e Ingenier\'ia, Universidad Aut\'onoma Metropolitana-Iztapalapa, San Rafael Atlixco 186, Col. Vicentina, Iztapalapa C. P. 09340, Ciudad de M\'exico, M\'exico.}
\address[c]{Institute for Advanced Computational Science, Stony Brook University, Stony Brook, NY 11749, USA.}
\address[d]{Instituto de Ciencias Físicas, Universidad Nacional Autónoma de México, Ap. Postal 43-8, Cuernavaca, Morelos, 62251, Mexico.}
\address[e]{Theoretische Chemie, Physikalisch-Chemisches Institut,
Universit\"at Heidelberg, INF 229, 69120 Heidelberg, Germany}

%%%%%%%%%%%%%%%%%%%%%%%%%%%%%%%%%%%%%%%%%%%%%%%%%%%%%%%%%%%%%%%%
%%%%%%%%%%%%%%%%%%%%%%%%%%%%%%%%%%%%%%%%%%%%%%%%%%%%%%%%%%%%%%%%
\begin{abstract}
We study the formation of C$_{18}$H and C$_{18}$H$_2$ by 
irradiating a cyclo[$18$]carbon molecule with atomic and 
molecular hydrogen at impact energy, $E$, in the range of
0.5-25 eV. 
We utilize the density-functional tight-binding method to 
perform molecular dynamics simulations to emulate the 
interaction of a carbon ring when colliding with atomic or 
molecular hydrogen. 
From our results, the formation of the C$_{18}$H molecules 
is likely to occur upon irradiating by H atoms at $E < 10$ eV 
and by H$_2$ molecules at $2 < E < 15$ eV center of mass energy. 
Formation of C$_{18}$H$_2$ molecules is only observed at 
around $E = 2$ eV. 
Our results show that the absorption of hydrogen is more 
prone in atomic than in molecular hydrogen atmosphere. 
Thus, we find that the probability of physio-absorption 
reaches up to 80 \% for atomic projectiles with $E < 5$ eV but 
only up to 10 \% for the molecular ones. 
Our analysis shows that the deformation of the carbon ring 
due to the hydrogen bonding produces transition from $sp$ to
$sp^2$ hybridization.  
The angle between the carbon atoms at the locations near to 
the H bond in the resulting ring is not 120$^o$ but instead 
110$^o$ degrees.  
No molecular fragmentation of the C$_{18}$ ring is observed.
\end{abstract}

\end{frontmatter}
%\noindent{\it Keywords}: cyclo18carbon, graphene, QCMD \\
%\submitto{\JPD}
%\maketitle
%\linenumbers
%%%%%%%%%%%%%%%%%%%%%%%%%%%%%%%%%%%%%%%%%%%%%%%%%%%%%%%%%%%
\section{Introduction}
\label{sec:intro}

The electronic structure of carbon atoms produces various 
allotropic forms. 
These forms include, for example, ball shapes of 
buckminsterfullerene, nanotubes, nanobuds, nanoribbons 
and 2D structures such as graphene. 
More unusual forms of carbon exist, one of them being 
just recently reported by Kaiser et al.
\cite{Kaiser-K19-365science1299}.
This allotrope is a ring of eighteen carbon atoms, named
cyclo[18]carbon or C$_{18}$. 
The C atoms are connected by alternating triple and single bonds, 
forming a polyyne and a cyclocarbon. 
The various allotropic forms are a distinctive feature of 
the $sp$-hybridization of carbon valence electrons. 
H\"uckel's rule predicts an aromatic structure with no bond 
length alternation for planar, cyclically conjugated systems 
with $(4m + 2)$ $\pi$-electrons, where $m$ is a natural number  
related to C$_N$ carbon rings with $N =4m+2$ as the number 
of C atoms \cite{Fowler-PW09-15chem6964}. 
Hoffmann predicted double aromatic stabilization consequence 
of two orthogonal ring currents in C$_{18}$ 
\cite{Hoffmann-R66-22tetrahedron521}.
Thus, a theoretical debate started on the chemical structure 
of cyclo[$N$]carbons with conclusions which have so far depended 
on the theoretical approach. 
Most of the density functional theory (DFT) and Møller-Plesset
perturbation theory calculations predict that the lowest-energy
geometry of C$_{18}$ is cumulenic D$_{18h}$
\cite{Parasuj-V91-113jacs1049,Neiss-Ch14-15cpc2497}. 
Hartree-Fock\textcolor{orange}{,} Quantum Monte 
Carlo, and coupled cluster methods predict that the polyyne 
D$_{9h}$ form is the ground state \cite{Diederich-F89-245science1088,Torelli-T0085prl1702,Arulmozhiraja-S08-128jcp114301,Pereira-S20-0jpca0,Baryshnikov-GV19-10jpcl6701}.
Recently, it has been found that C$_{18}$ is the smallest among 
carbon electron acceptor molecule reported so far
\cite{Stasyuk-A20-56cc352} and its synthesis
\cite{Kaiser-K19-365science1299} paves the way for the creation 
of various 2D materials based on molecular carbon allotropes, 
e.g. it might be used  to produce graphdiyne (GDY)
\cite{Diederich-F94-369nature199,doi:10.1002/andp.201700056}.
GDY is of interest for atom catalyst where the hydrogen evolution
reaction at room temperature is observed during its synthesis 
with zero-valent Mo atoms \cite{doi:10.1021/jacs.9b03004}.
This catalytic process can be applied for hydrogen production 
with application on energy storage. The material has a porous
structure that allows adsorption and diffusion of atoms, a highly
desirable feature for the design of next-generation batteries. 
Because of the presence of a natural band gap and high degree of
conjugation, GDY is also a promising material for nanoelectronics
applications. 
Therefore, the study of the formation of C$_{18}$ molecules, with or
without hydrogen atom or molecules, is needed to understand the
$sp$-$sp^2$-hybridized carbon atoms formed by inserting 
diacetylenic (C$_{4}$H$_{2}$) linkages between two benzene rings
in a graphene structure  \cite{Baughman-RH87-87jcp6687}.
The formation of the C$_{18}$H$_n$ $(n = 1, 2, \ldots, 36)$ 
molecules leads to anisotropy in the optical activity of the 
original C$_{18}$ molecule, where the molecular geometry changes 
can be related to transition magnetic and electric dipole 
moments \cite{Gorter:ha0138}.

The previous discussion sets the stage and motivates our quantum
chemistry theoretical study: the absorption of hydrogen atoms by
a C$_{18}$ molecule irradiated by atomic and molecular hydrogen 
to provide an insight in the formation of C$_{18}$H and 
C$_{18}$H$_2$ and its molecular geometry modifications leading 
to a better understanding of this ring molecule in the development
of hydrogen nano sensors. 
This requires a computationally expensive methodology to obtain
results with chemical accuracy. 
The hydrogen uptake rates are highly dependent on the potential
barriers and adsorption energies requiring a statistical approach
to take into account all the possible projectile trajectories, 
which lead to the formation of a C$_{18}$H molecule. 
Consequently, the Density Functional Theory (DFT) method is an
ideal first candidate that provides accurate information about 
the electronic structure of the C$_{18}$ molecule and the 
$sp$- hybridization when the H atom is bound to a C atom in 
the C$_{18}$ ring. 
However, the H adsorption and scattering processes have a time 
scale that goes from femto to pico seconds
\cite{doi:10.1063/1.2899649}. 
This is a formidable computational 
task for the quantum-classical molecular dynamics
simulation based on DFT and  Carr-Parinello frameworks due of 
the need of a short time step of 0.05 fs and thousands of time 
steps to model the interaction between a carbon ring with an 
atomic or molecular projectile,which is needed to track the formation
of C$_n$H$_m$ molecules with n and m as integer numbers.
Instead, in this work, we use the less expensive semi-classical
Molecular Dynamics (MD) approach based on the computationally
affordable density-functional tight-binding (DFTB) method 
\cite{doi:10.1021/jp070186p,PhysRevB.58.7260}, as implemented 
in the dftb$+$ code \cite{dftb}. 
This method has the capability to compute energy contribution
due to large interatomic distances, spin effects, electron-electron 
repulsion terms, and 3-body dispersion correction with a numerical 
accuracy close to those obtained by DFT simulations. 
Thus, we perform MD simulations in the impact energy range of 0.5 to 
25 eV for 100 fs to estimate the probability of formation of a
C$_{18}$H$_X$ molecule and to analyze the mechanisms of H capture
by the carbon ring. 
This analysis could elucidate the feasibility for designing a 
fast and sensitive hydrogen nano sensor device. 

Our paper is organized as follows: In Sec. \ref{sec:com_meth},
we briefly discuss the numerical method used in our work, as 
well as computation of the H-C$_{18}$ potential energy curves. 
In Sec. \ref{sec:Results}, we report the results of our 
simulations. 
Our concluding remarks are given in Section \ref{sec:Concl.}.

%%%%%%%%%%%%%%%%%%%%%%%%%%%%%%%%%%%%%%%%%%%%%%%%%%%%%%%%%%%%
\section{Theoretical approaches}
\label{sec:com_meth}

\subsection{SCC-DFTB method}

DFTB is approximately 100 times more expensive than the classical force fields and up to 100 times cheaper than density functional theory. Thus, it fills the gap between classical force fields and density functional theory and is an attractive candidate for direct molecular dynamics simulations of bulk phase and condensed matter \cite{doi:10.1021/ct300849w,doi:10.1021/jp070186p,Elstner-M01-263cp2,PhysRevB.58.7260}. 
DFTB is an approximate density functional theory in which only 
valence electrons are treated quantum mechanically while all core
electrons and nuclei are approximated via pairwise interatomic
repulsive potential 

\begin{equation} \label{eq1}
\begin{split}
E & = \sum_i 2f_i \langle \phi_i | H_{core} | \phi_i \rangle + 
\frac{1}{2}S \sum_{\substack{A,B \\ A\neq B}} \gamma^{AB}\Delta q^A \Delta q^B \\
& + \sum_{A>B}^{Atom}E_{rep}^{AB} 
+\frac{1}{2} S
\left( \sum_{l' \in A} W_{All'}m_{Al'} 
+ \sum_{l'' \in B} W_{Bl'l''}m_{Bl''} \right)
\end{split}
\end{equation}

Here $f_i$  is an occupation number (typically 0 or 1) and $i$  
runs over all molecular orbitals. 
The first term describes the interaction of valence electrons 
with core ions (nuclei and core electrons). 
The second term is responsible for electron-electron interaction. 
The symbols $\Delta q^A$  and  $\gamma^{AB}$ are, respectively, 
a charge at center $A$ and a chemical hardness-based parameter
describing electron-electron interactions between centers $A$ 
and $B$; 
$\gamma^{AB}$ depends on the interatomic distance; 
and the matrix $S$ is the overlap term.
The third term describes the interaction between core ions and is obtained 
from a fit. 
An important feature of DFTB that is often missing in standard 
DFT is a correct Coulomb asymptotic behavior for interaction of charged molecules \cite{doi:10.1021/ct300849w,PhysRevB.58.7260}. 
This is due to the fact that $\gamma^{AB}$, in the 
electron-electron repulsion term, behaves as $1/R_{AB}$ for 
large interatomic distances. 
The last term corresponds to the spin contribution and depends 
on the spin channel, $\sigma$. 
Here, the spin coupling constant $W_{All'}$ defines the spins
interaction where $l$ and $l'$ are shells on the same atom. 
The spin polarization $m_{Al'}$ is the difference between the 
Mulliken charges on spin up and down of the atom and is given 
as $m_{Al'} = q_{Al \uparrow}-q_{Al' \downarrow}$
\cite{doi:10.1021/jp070186p}. 
DFTB provides also inexpensive tools for the description of 
lowlying excitations. 
Higher energy electronic excitations are less reliable. 
This limitation is a result of inherently minimal basis set 
(Slater type orbitals).  
Higher energy excitations can be included through all electrons
DFT approaches and extended basis set. The DFTB approach, as
implemented in the publicly available dftb+ code \cite{doi:10.1021/jp070186p,dftb} version 19.1, is used in 
this work. 
Molecular Dynamics simulations are performed by using the DFTB 
method with a velocity-Verlet algorithm \cite{doi:10.1021/jp070186p,PhysRevB.58.7260,doi:10.1021/ct300849w,Elstner-M01-263cp2}.

\subsection{Potential energy curves}
\label{subsec:pecs}

As a first step in our study, we calculate the adiabatic potential 
energy curves of H and H2 interacting with C$_{18}$. 
The absorption of H is strongly dependent on the electronic 
structure of the C$_{18}$H system.  
We need a correct set of Slater-Koster (SK) parameters. 
For this purpose, we tested various SK parameter sets from 
the literature \cite{doi:10.1021/jp070186p,Lukose,RAULS1999459,VuongVQ} implemented in 
the DFTB+ code, in order to select the most suitable
one for our systems. With appropriate well-chosen SK parameters, 
one should be able to model the $s-p$ hybridization in the C$_{18}$H 
in good agreement with DFT calculations. 
Our first choice is the semi-relativistic, self-consistent charge
set of SK parameters for materials science (MATSCI), which 
was successfully used to study the formation of 2D Covalent 
Organic Frameworks \cite{Lukose}. 
We utilized these parameters in our previous work for the study 
of electronic properties of a borophene sheet \cite{novotny} 
and hydrogen uptake by carbon, boron doped carbon fullerenes 
\cite{DOMINGUEZGUTIERREZ2018189}, and C/B-N nanotubes
\cite{doi:10.1063/1.5079495}. 
Our second choice is the set of parameters for solid-state 
systems (PBC) \cite{RAULS1999459} that has been used to study 
SiC properties and could be applied to organic systems. 
As a third option, we consider the long-range corrected SK 
parameters for bio and organic molecules \cite{ob2-1-1} (OB2), 
which have been used in our previous work for the study of the
effects of hydrogen radiation on a glycine molecule
\cite{MARTINEZFLORES2020108513}. 
In order to explore the advantages of the inclusion of the 
third order corrections of the DFT energy expansion into the 
DFTB approach \cite{doi:10.1021/ct300849w}, we also use the SK parameters called 3OB \cite{VuongVQ} that are focused on the 
study of small organic molecules. 
The choice of these parameters reflects the fact that the 3OB 
were generated to improve hydrogen binding energies, proton
affinities, and proton transfer barriers. 
To include van der Waals interactions between H and C in our
computations, we use the following three different levels of 
theory for the dispersion corrections: the Slater-Kirwood 
polarizable atomic model (SKirk) \cite{PhysRev.37.682}, 
the Lennard-Jones potential  with parameters taken from the 
Universal Force-Field (UFF) \cite{UFF_potentials}, and the 
3-body term, included through a damped pairwise London-type
correction (DFTD3) \cite{doi:10.1021/acs.jctc.7b00118}. 

In Table \ref{tab:tab1}, the bond lengths of the ground state
geometry configuration of C$_{18}$ are given in units of \AA{}, as
obtained by DFTB in good agreement with the DFT results by Kaiser 
et al \cite{Kaiser-K19-365science1299}. 
As a consequence of the $sp$ hybridization, we find that the 
C$_{18}$ molecule has an alternating bond length resulting of 
single and triple bonds between the carbon atoms forming the ring. 
We report these alternating bonding lengths by considering a 
system of three carbon atoms C(2)-C(1)-C(3). We find that in 
general all the SK parameters, i.e. MATSCI, PBC, OB2, and 3OB,
produce the same geometry of the ring molecule. 
This confirms the D$_{9h}$ symmetry of this molecule 
\cite{Torelli-T0085prl1702,Arulmozhiraja-S08-128jcp114301,Pereira-S20-0jpca0,Baryshnikov-GV19-10jpcl6701} for all cases. 

\begin{table}[!t]
\caption{Bond lengths of the C$_{18}$ molecule obtained after an optimization procedure by the DFTB methods and are compared to the DFT data from Kaiser et al. \cite{Kaiser-K19-365science1299}
We report values by using MATSCI, PBC, OB2, and 3OB parameters for the DFTB calculations.
\label{tab:tab1}
}
\centering
\begin{tabular}{l c c c c | c }
\hline
        & \multicolumn{4}{c }{Length (\AA)}  \\ 
\hline
Bond        & MATSCI & PBC & OB2 & 3OB &  DFT$^{[1]}$\\
\hline
C(1)--C(2)  & $1.25$ & $1.25$ & $1.27$ & 1.25 &  1.20  \\
C(1)--C(3)  & $1.35$ & $1.35$ & $1.35$ & 1.35 &  1.34  \\
\hline
\end{tabular}
\end{table}

%%%%%%%%%%%%%%%%%%%%%%%%%%%%%%%%%%%%%%%%%%%%%%%%%%%%%%%%%%%%%%%%%%%%%%%%%%%%%%%%%%%%%%

In order to calculate the potential energy curves (PECs) of a 
C$_{18}$H and C$_{18}$H$_2$, the systems consisting of C$_{18}$ 
ring and H or H$_2$ are first considered. 
The PECs are calculated as $\Delta E(r) = E_{HC}(r)-E_H -E_C$, 
where $E_{HC}$ is the total electronic energy of the 
C$_{18}$H$_X$ system. 
E$_H$ is the energy of the isolated H or H$_2$ and $E_C$ is 
the energy of the isolated C$_{18}$ ring. 
All the atomic nuclei are fixed during the energy calculations 
while the radial directions are chosen with respect to 
different adsorbate sites. 
We consider absorbate sites that determine the D$_{18h}$ or 
D$_{9h}$ symmetry axis of the ring where the H atom or H$_2$ 
molecule is placed at several positions on a straight line 
parallel to the x-y plane starting at the center of the 
C$_{18}$ molecule on the direction going through the 
C(1)-C(2), C(1)-C(3) bonds, in the radial directions toward 
the center of the ring, and in the plane of the ring and the 
C atoms location. 
We compute the PECs with the DFTB$+$ code considering the 
MATSCI, PBC, OB2, and 3OB SK parameters with UFF, 
Slater-Kirwood, and DFTD3 dispersion corrections, which help 
us to select the SK parameters set to perform MD simulations.

\subsection{Irradiation on C$_{18}$ by Atomic and Molecular Hydrogen }
\label{subsec:irradiation}

\begin{figure}[!t]
   \centering
   \includegraphics[width=0.48\textwidth]{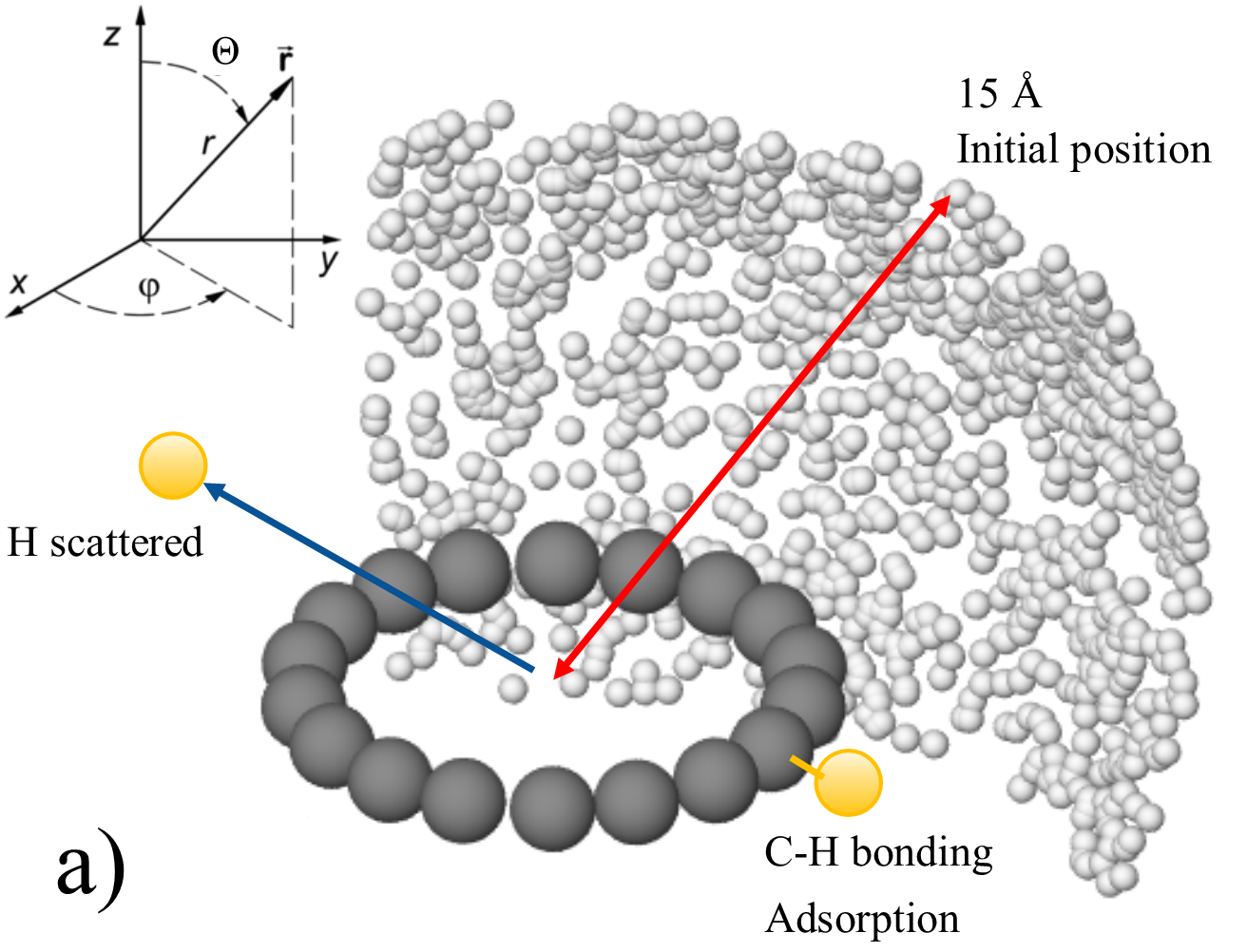}
   \includegraphics[width=0.48\textwidth]{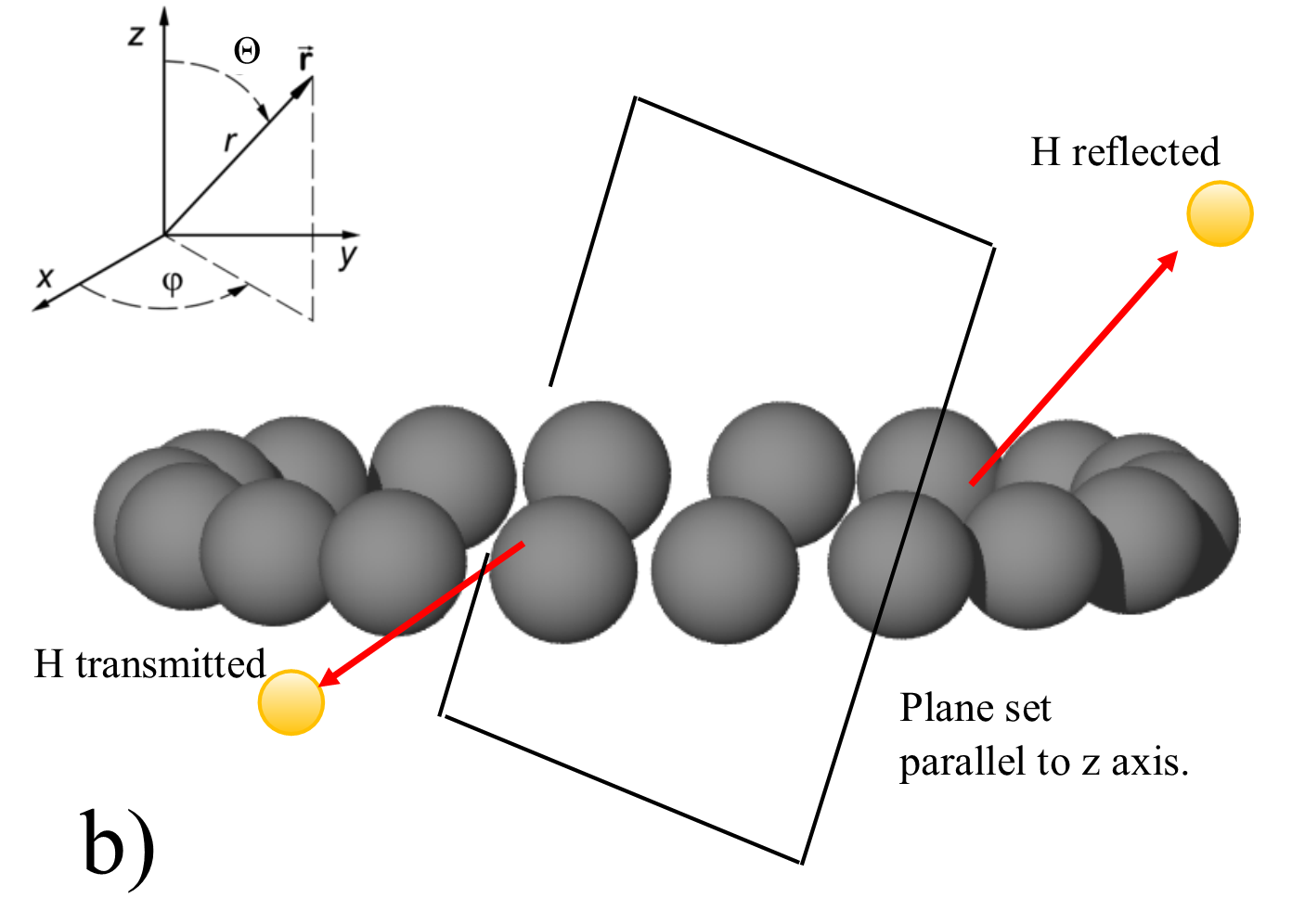}
   \caption{(Color on-line). Schematic view of the initial 
   positions for H$_2$ projectiles with random orientations 
   and the C$_{18}$ molecule in a). 
   The carbon ring is set at the origin of the coordinate 
   system, in the x-y plane. 
   We also include a representation of the scattered and 
   adsorbed H atom by showing them as yellow spheres. 
   In Fig. \ref{fig:fig1} b), we show the interpretation of 
   the transmitted and reflected H atoms by introducing a 
   plane that divides the incident projectiles from the 
   scattered region that is vertical along the z axis and 
   perpendicular to the plane formed by the C$_{18}$ molecule, 
   which determines the number of transmitted and reflected atoms. 
   Color code: H atoms are illustrated as white spheres, 
   meanwhile C atoms are presented by gray spheres.  }
   \label{fig:fig1}
\end{figure}

The study of the formation of a C$_{18}$H or C$_{18}$H$_2$ 
molecule is carried out by performing molecular dynamics 
simulations using the DFTB approach. 
The center of the carbon ring is placed at the origin of the
Cartesian coordinate system, then the ring is energy optimized
and thermalized to 300 K using a Nose-Hoover thermostat 
\cite{doi:10.1080/00268979600100761} prior to hydrogen irradiation.
%The choice of the 5 K temperature is to emulate, as close as
%possible, the experimental conditions under which the C$_{18}$
%molecule has been synthesized.  
%Then the system C$_{18}$ is thermalized to 300 K prior 
%bombardment \cite{Kaiser-K19-365science1299}.
%\textcolor{blue}{Then, what was the porpuse of the 5 K thermalization? The referee may ask whay we did it!}
We consider the following impact energies: 
0.5, 1, 2, 5, 10, 15, 20, and 25 eV with a commensurate 
velocity given by $v = \sqrt{(2E/m)2}$, where $E$ is the 
impact energy and $m$ is the mass of the projectile. 
We irradiate the C$_{18}$ molecule with 942 hydrogen atoms for 
the atomic case and with 471 H$_2$ molecules for the molecular 
case. 
The initial H and H$_2$ positions start at a distance of 15 \AA{}
measured from the origin of our coordinate system and are randomly
distributed in a target area of 1 nm$^2$ considering a spherical
geometry \cite{DOMINGUEZGUTIERREZ2018189} with the azimuth angle
being in a range of $0 \leq \varphi \leq \pi/2$ and the zenith angle,
defined in the range $–\pi/2 \leq \Theta \leq \pi/2$. 
The irradiation occurs atom per atom (or molecule) rather than 
in a cumulative process and a thermostat is not used during the 
MD simulations for collision dynamics. 
In Fig. \ref{fig:fig1} a), we show the initial positions of the 
H$_2$ molecules (represented by white spheres) and the C$_{18}$
molecule (depicted as gray spheres) used in this work. 
Each H$_2$ molecule has initially a random orientation. 
For the collision dynamics, we use the velocity-Verlet algorithm 
as implemented in the dftb+ package with a time step of 
$\Delta t=0.05$ fs with the total simulation time of 100 fs. 
A Fermi-Dirac smearing is applied in the MD simulations with
equivalent electronic temperature of 1000 K. 

Once the carbon ring and the atomic or the molecular hydrogen
projectiles are prepared, the MD simulations are performed with
the dftb$+$ code. 
The analysis of the MD simulations is done by quantifying the 
number of H atoms $N^{a,s}$, that are absorbed, $a$, or scattered,
$s$, by the carbon ring. 
From the output file of our simulations, we compute the distance
$d_f = \sqrt{x^2+y^2+z^2 }$ from the final position of each H 
atom, defined by $x$, $y$, and $z$, to the center of the carbon 
ring. 
Scattered and adsorbed H atoms are shown in Fig. \ref{fig:fig1} 
a) by yellow spheres. 
Bound atoms are commonly found with $d_f  < 7$ \AA{}, and 
scattered atoms have a $d_f \gg 7$ \AA{}.  
The adsorption and scattered probabilities are calculated as 
$N^{a,s} /N_T$ for each impact energy with $N_T$ defined as 
the total number of impinging projectiles. 
The scattering particles can be divided into transmitted 
and reflected projectiles. 
Thus, we compute the angle between the associated vectors to 
the initial $\vec r_i$ and final $\vec r_f$ positions of the H 
atom as angle $\theta_t = \arccos{\left( \hat r_i \cdot \hat r_{\textcolor{red}{f}} \right)}$, with 
$\hat r_{i,\textcolor{red}{f}}$ normalized vectors.
The final commensurate velocity of the projectile is computed 
from the output data of the MD simulations as 
$v_f =\sqrt{v_x^2+v_y^2+v_z^2 }$ where $v_i$ is the final 
velocity component in the x, y, and z directions. 
In this way, H atoms with $\pi/2 < \theta_t < 3\pi/2$ and a
$v_f$ similar to the initial velocity are considered as 
transmitted, otherwise the projectiles are identified as 
reflected, as shown in Fig. \ref{fig:fig1} b). 
The probability of these processes is calculated as $N^{β,t} /N_{st}$ 
where $N^{β,t}$ is the number of reflected ($\beta$) or 
transmitted ($t$) H atoms and $N_{st}$ is the total number 
of scattered H atoms. 
This ratio provides an insight of the probability of a 
C$_{18}$H or C$_{18}$H$_2$ molecule formation due to 
hydrogen irradiation as a function of the irradiation energy.

%%%%%%%%%%%%%%%%%%%%%%%%%%%%%%%%%%%%%%%%%%%%%%%%%%%%%%%%%%%%
\section{Results}
\label{sec:Results}

\subsection{Static description}
\subsubsection{PECs: Atomic case}
\label{subsubsec:atomic}

\begin{figure}[!t]
   \centering
   \includegraphics[width=0.9\textwidth]{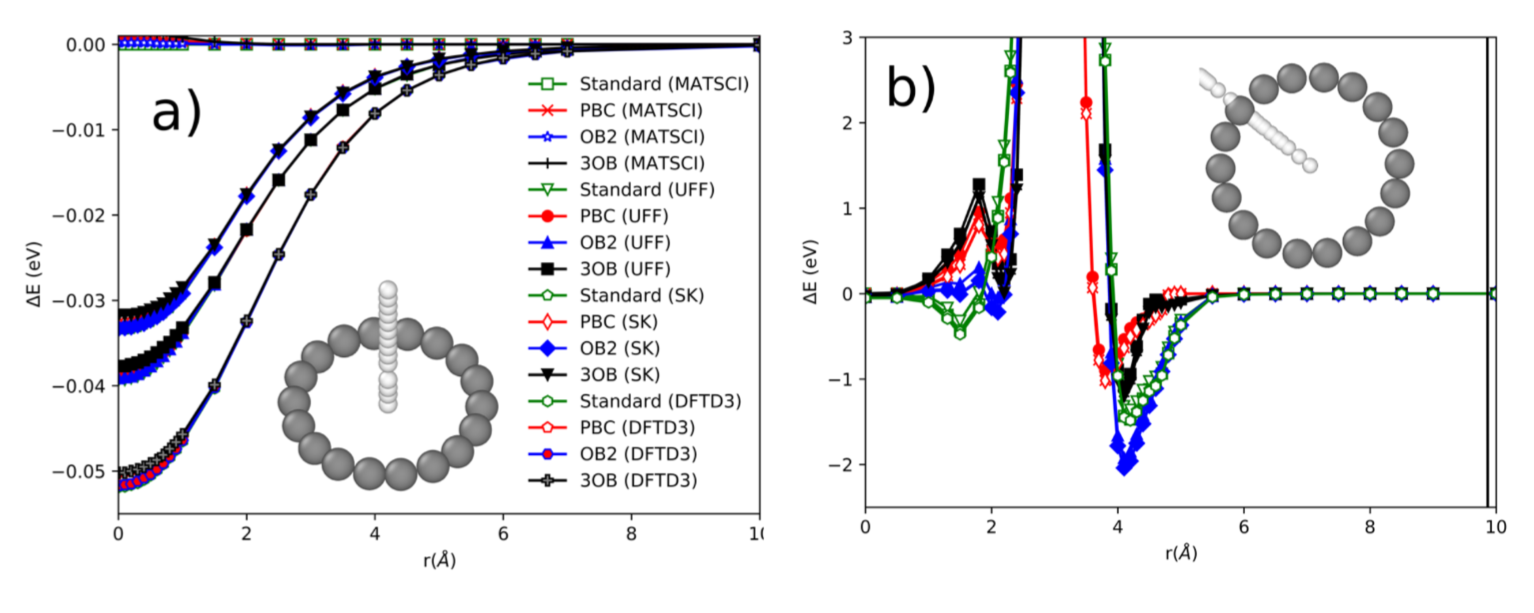}
   \caption{(Color on-line). a) Potential energy curves for the formation of a C$_{18}$H molecule for several SK parameters including dispersion corrections. 
   A schematic of the positions  of the hydrogen atom and the 
   carbon ring is shown as an inset. 
   a) The hydrogen atom is placed along the $D_{9h}$ symmetry axis. 
   b) The hydrogen atom is placed on the plane of the C$_{18}$ 
   molecule and passes through a C carbon atom. (color code: 
   white spheres represent hydrogen atoms and carbon are depicted
   as gray spheres.) See text for discussion.}
   \label{fig:fig2}
\end{figure}

\begin{figure}[!b]
   \centering
   \includegraphics[width=0.9\textwidth]{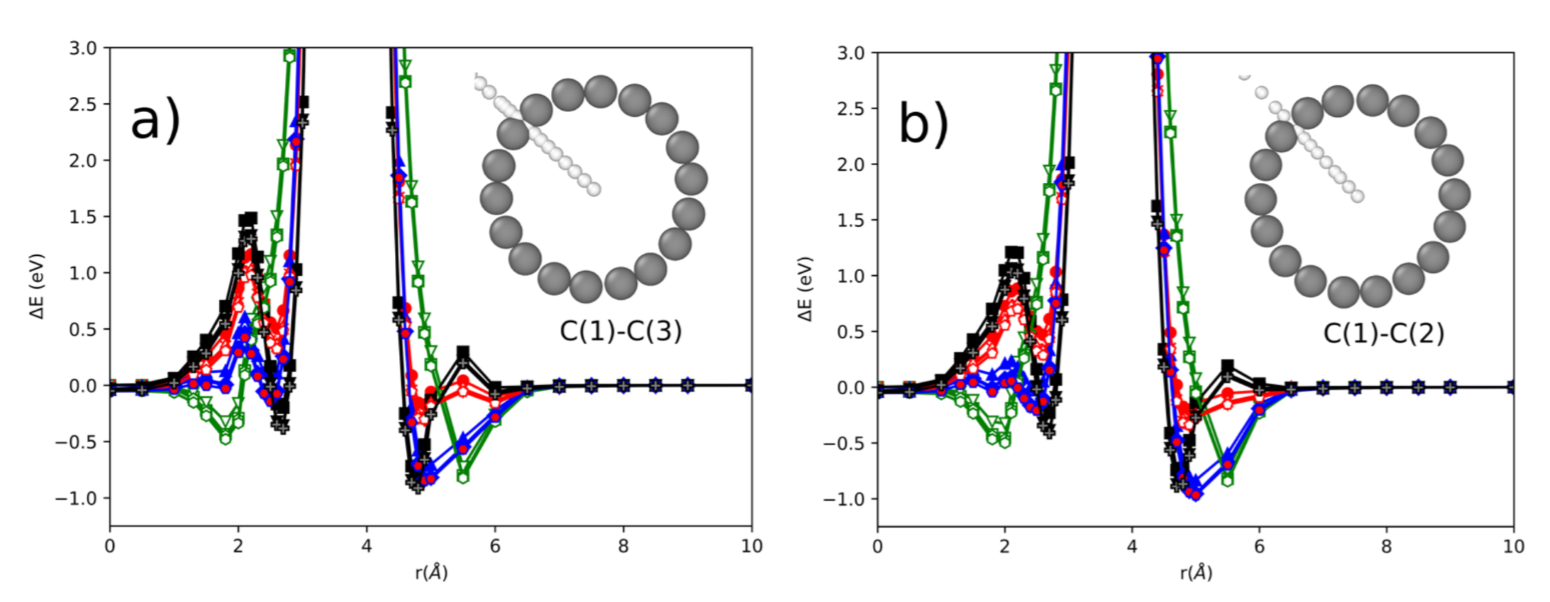}
   \caption{(Color on-line). Potential energy curves for C$_{18}$ 
   and a H atom as a function of the distance. 
   Here, we consider two different adsorbate sites, which are 
   shown schematically as an inset in the figure. 
   Different SK parameters and dispersion approaches are 
   tested. 
   Color code and labels are the same as in Fig. \ref{fig:fig2}}
   \label{fig:fig3}
\end{figure}

In Figs. \ref{fig:fig2} and \ref{fig:fig3}, we present the 
PECs obtained for a hydrogen atom and a carbon ring molecule 
placed at a distance $r$ from the center of the ring by 
considering different SK parameters that describe the electronic
structure of C$_{18}$ and its interaction with a H atom. 
In Fig. \ref{fig:fig2} a), the hydrogen atom is displaced
perpendicular to the plane of the C$_{18}$ molecule passing 
through the center of the ring. The standard results, obtained
without dispersion corrections, show that the hydrogen atom cannot
be bound to the carbon ring at this adsorbate configuration. 
The inclusion of dispersion corrections in the computation of 
the PECs improves the bonding energy to 0.039 eV for the Slater
Kirwood approach, 0.032 eV for the UFF, and 0.051 eV with the 
DFTD3. 
This is still a too small binding energy to bind H at the center of
the ring. 
In Fig. \ref{fig:fig2} b), H atoms are placed along a parallel 
line on the plane of the carbon ring, going through a C atom. 
This geometry shows different results when varying the SK 
parameters.  
MATSCI and OB2 parameters bind a hydrogen atom outside the 
carbon ring, at a distance of 4 \AA{} from the origin of the ring
or 0.6 Å from the carbon atom.
We find that the OB2 parameters, which are obtained for 
biomolecules in gas phase, show the largest binding distance and
energy. 
Since the SK parameters are obtained from different training 
data sets or reference molecular systems, the results vary for 
the bond length and binding energy. 
MATSCI and PBC parameters are obtained for modeling a solid-state
system and cannot model properly our present cases.  
The DFTD3 includes the spin-same-spin and spin-different-spin
constants for all combinations of atomic shells and one-center
exchange-like terms for the multicenter integrals
\cite{doi:10.1021/acs.jpca.5b01732}. 
In Fig. \ref{fig:fig2} b), we observe that the H atom is bound 
to the C$_{18}$ with an energy of 1 eV, which is the result of the
inclusion of the DFTD3 dispersion corrections, where the total
energy of the system is expressed as $E=E_{DFTB}+E^{(2)}_{disp}+E^{(3)}_{disp}$,
considering 2- and 3-body contributions in the calculations. 
Our results indicate that the 3OB SK parameters are a good choice
to perform MD simulations to emulate the interaction of the 
carbon ring with the hydrogen projectile.

In Fig. \ref{fig:fig3}, we show results for the PECs for the two
adsorbate sites, consisting of the hydrogen atom passing through
the bridge between the C(1)-C(2) and C(1)–C(3) bonds according 
to the $D_{9h}$ symmetry of the carbon ring. 
In Fig. \ref{fig:fig3} a), we present the results when the H atom
is placed between the line that joins the C(1)-C(2) atoms. 
We observe that the different SK parameter sets show differences
in the PECs, but the inclusion of dispersion corrections does not
affect the results. 
The PBC SK parameters show a repulsion of the hydrogen atom in 
the inner part of the carbon ring, consequently, the formation 
of C$_{18}$H at the inside ring is not possible with these parameters.
MATSCI SK parameters model the formation of a C$_{18}$H with a 
hydrogen bonding inside and outside the carbon ring. 
In addition, the OB2 parameters model a C$_{18}$H molecule with
a hydrogen atom bound outside of the ring. 
However, the repulsion region inside the carbon ring is similar
to that obtained by the PBC SK parameters. 
With the 3OB SK parameters, we get that H bounds more favorable
outside of the carbon ring. 
Although the repulsion region inside the C$_{18}$ is similar to
the results presented by OB2 and PBC, the H atom can be bounded
to the carbon ring in this region. 

In Fig. \ref{fig:fig3} b), we present results for the adsorbate
site defined in the bridge between the C(1)-C(2) carbon atoms. 
The results are similar to those reported in Fig. \ref{fig:fig3}
a), and the hydrogen atoms can be bounded to the carbon ring at
the outside region. 
The PBC SK parameters show slight differences when comparing 
the C(1)-(C3) to the C(1)-C(2) bond which, in principle, reflects
the proper binding of a H to the C$_{18}$ molecule resulting from
a $D_{9h}$ target symmetry. 
In summary, the H atom is more likely to be bound to a C atom to
form a C$_{18}$H molecule. 

%%%%%%%%%%%%%%%%%%%%%%%%%%%%%%%%%%%%%%%%%%%%%%%%%%%%%%%%%%%%%%%%%%%%%%%%%%%%%%%%%%%%%%%%%%%%%%%%%%%%%%%%%%%%%%
\subsubsection{PECs: Molecular case}
\label{subsubsec:molecular}

\begin{figure*}[!b]
   \centering
   \includegraphics[width=0.90\textwidth]{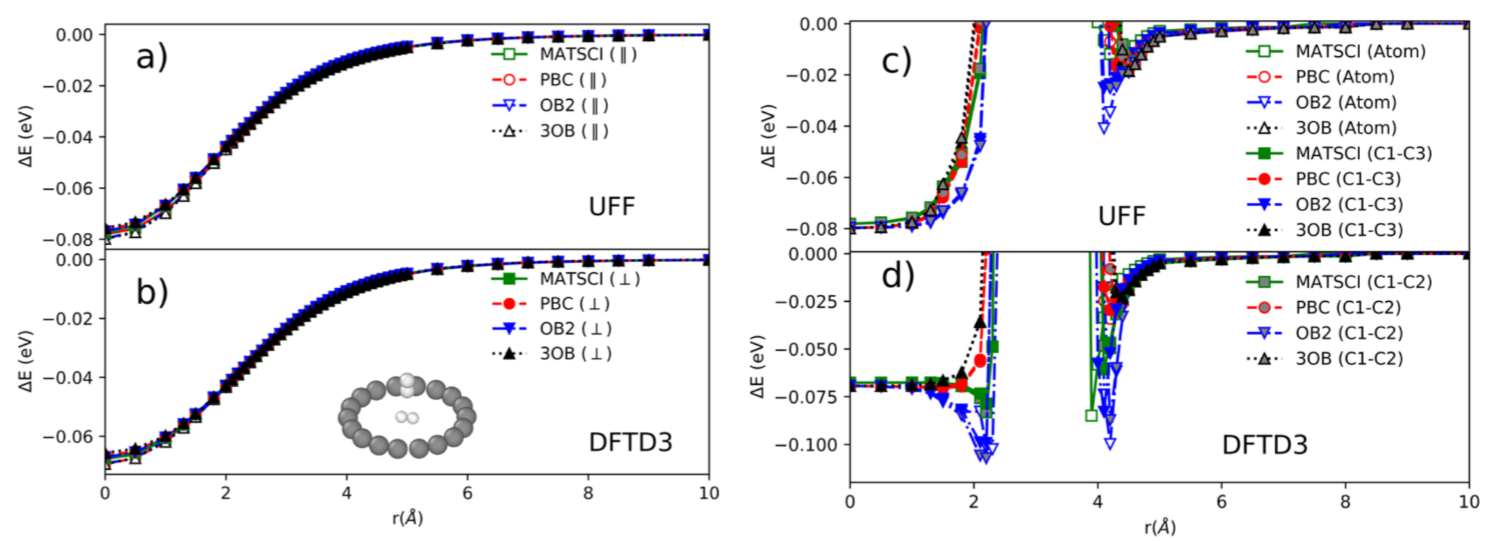}
   \caption{(Color on-line). Potential energy curves for the
   interaction of a C$_{18}$ with a H$_2$ molecule placed along 
   the symmetry axis perpendicular to the ring plane and at 
   the center of the carbon ring in a) and by considering 
   three different adsorbate sites in b). 
   The positions of the H$_2$ molecule center of mass follow 
   straight lines to the different adsorbates site as shown 
   in Fig \ref{fig:fig2} and \ref{fig:fig3}. 
   Hydrogen-hydrogen repulsion corrections are considered in 
   the computation of the PECs. 
   Similar results are obtained for perpendicular and parallel
   orientation of the H$_2$ molecule respect to the plane formed
   by the carbon ring. 
   See text for discussion.}
   \label{fig:fig4}
\end{figure*}

In Fig. \ref{fig:fig4}, we compute the PECs of the interaction 
of a H$_2$ molecule at different adsorbate sites of the C$_{18}$
molecule by considering two orientations of the H$_2$ molecule, 
one horizontal and the other vertical with respect to the plane
of the ring molecule. 
The center of mass of the hydrogen molecule is set at different
locations along a line that goes through the three absorbate sites
of the carbon ring. 
In Fig. \ref{fig:fig4} a), we present the PEC results by using the same
MATSCI, PBC, OB2, and 3OB SK parameters with UFF and DFTD3 
dispersion corrections for a H$_2$ where the center of mass of the 
hydrogen molecule follows a line
perpendicular to the plane formed by the carbon ring. 
The inset figure shows two different orientation of the H$_2$ 
molecule, where white spheres represent a hydrogen molecule with
an orientation parallel ($\parallel$) to the carbon ring plane, 
while silver spheres show a H$_2$ molecule perpendicular ($\perp$) 
to the C$_{18}$ molecule plane. 
Hydrogen bonding corrections \cite{doi:10.1021/ct200751e} are
included during the computation of the PECs. 
Both orientations of the hydrogen molecule present a bonding 
energy of 0.08 eV for the UFF dispersion, while an energy bonding
of 0.07 eV is shown by the DFTD3 results in the PEC. 
Both orientations of the hydrogen molecule are favorable for 
binding H$_2$ to the C$_{18}$ molecule at this adsorbate site.

In Fig. \ref{fig:fig4} b), PECs for three different adsorbate 
sites are presented. 
We do not find significant differences between the results for
the two orientations of the H$_2$ molecule. 
Thus, we present only results for the perpendicular orientation
case. 
The results obtained by using UFF dispersion corrections show 
a small energy bonding of 0.02 eV outside of the carbon ring 
(at a distance of 0.6 \AA{} from the C atom) for most of the SK 
utilized in the calculations. 
The lower panel of Fig. \ref{fig:fig4} b) shows results for the
DFTD3 dispersion corrections and hydrogen-hydrogen repulsion
contributions. 
MATSCI and OB2 SK parameters results are affected by the inclusion
of DFTD3, as shown by the comparison to the results obtained with
UFF corrections. 
Also, MATSCI and OB2 SK parameters yield that the H$_2$ molecule 
can be bound at both sides of the carbon ring. 
The 3OB SK parameters results with DFTB3 dispersion corrections 
show similar behavior with those obtained by UFF. 
The energy bonding is small and the H$_2$ molecule is not 
attracted to the carbon ring. 
Consequently, the formation of a C$_{18}$H molecule can be 
feasible by dissociating the H$_2$ molecule through the 
collision dynamics (see below). 

Consequently, the results presented for the OB2 and 3OB SK 
parameters are suitable to model the interaction of the C$_{18}$
molecule with molecular hydrogen and will be used to perform 
the QCMD simulations. 
The later includes DFTD3 dispersion corrections 
(hydrogen-hydrogen repulsion contributions).

\subsection{Dynamical description}
\subsubsection{Atomic H projectiles}
\label{subsec:c18hformation}

\begin{figure}[!b]
   \centering
   \includegraphics[width=0.48\textwidth]{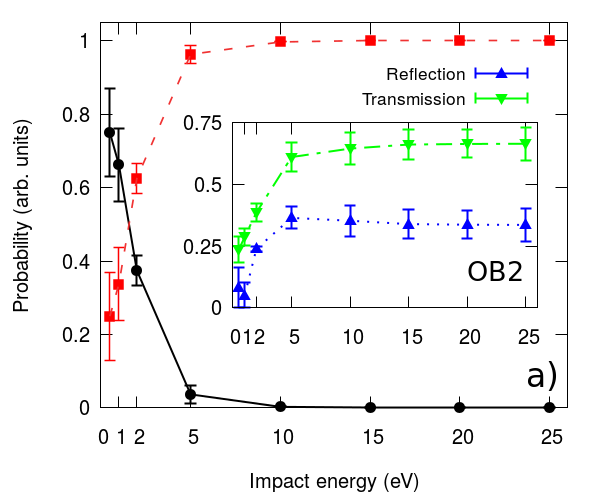}
   \includegraphics[width=0.48\textwidth]{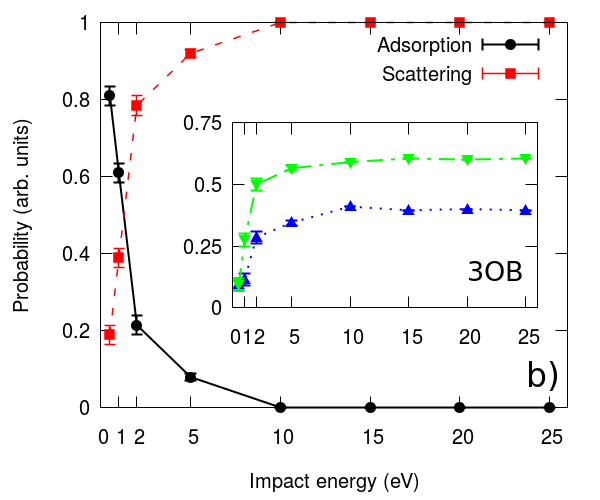}
   \caption{(Color on-line). Probability of adsorption and 
   scattering of H projectiles as a function of the impact energy
   when irradiating a C$_{18}$ molecule with the DFTD3 dispersion
   corrections for OB2 parameters in a) and for the 3OB SK ones 
   in b).  
   Probability of transmission and reflection of the H
   projectiles are shown in the inset graph. 
   Note that the probability of scattering $P_s = P_{\beta}+P_t$ 
   with $P_{\beta,t}$ is the probability of reflection ($\beta$) 
   or transmission ($t$) (See text for details). }
   \label{fig:fig5}
\end{figure}

In Fig. \ref{fig:fig5}, we report the probability of adsorption 
and scattering of a hydrogen atom as a function of the impact 
kinetic energy. 
Standard error bars, as obtained by the statistics of our 
sampling, are included in the results. 
The obtained results by the Slater-Kirwood, UFF, DFTD3 dispersion
corrections are similar for impact energies higher than 5 eV for 
OB2 (Fig. 5a) and 3OB (Fig. 5b) SK parameters. 
Note that the number of reflected and transmitted H atoms for 
impact energies higher than 5 eV is similar for both dispersion 
SK parameters. 
In this impact energy range, the interaction between the carbon 
ring and the H impact is negligible and the two methods report 
similar results. 
The OB2 MD simulations results show different behavior for the 
number of H atoms bound to the C ring due to the sp orbital
hybridization. 
The 3OB simulations results are more stable with respect to the 
dispersion corrections included in these calculations. 
However, an important difference is observed at 2 eV where 
40 \% of the H atoms are bound for the OB2 MD simulations while 
20 \% of them form a C$_{18}$H as obtained by the 3OB MD 
simulations. 
This can be understood from the PECs of these SK parameters 
(Figs. \ref{fig:fig2}-\ref{fig:fig3}), where the potential well 
at the outer region of the carbon ring is deeper or wider for 
the OB2 SK parameters than those reported by the 3OB SK ones at
the same interaction distances.

\begin{figure}[!t]
   \centering
   \includegraphics[width=220pt,height=200pt]{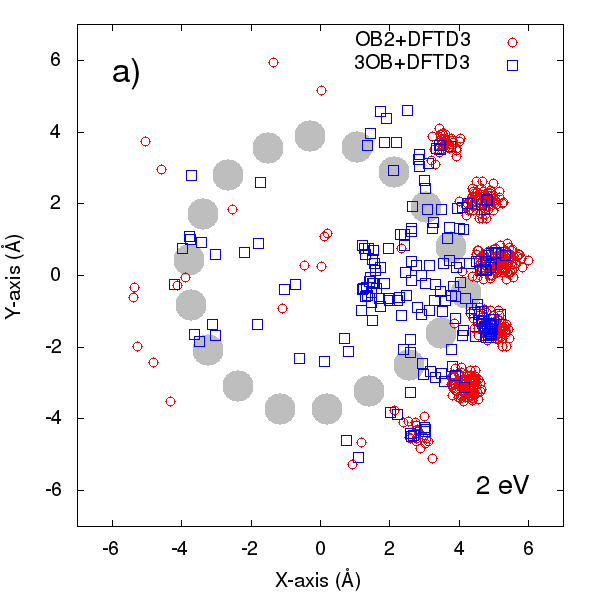}
   \includegraphics[width=230pt,height=220pt]{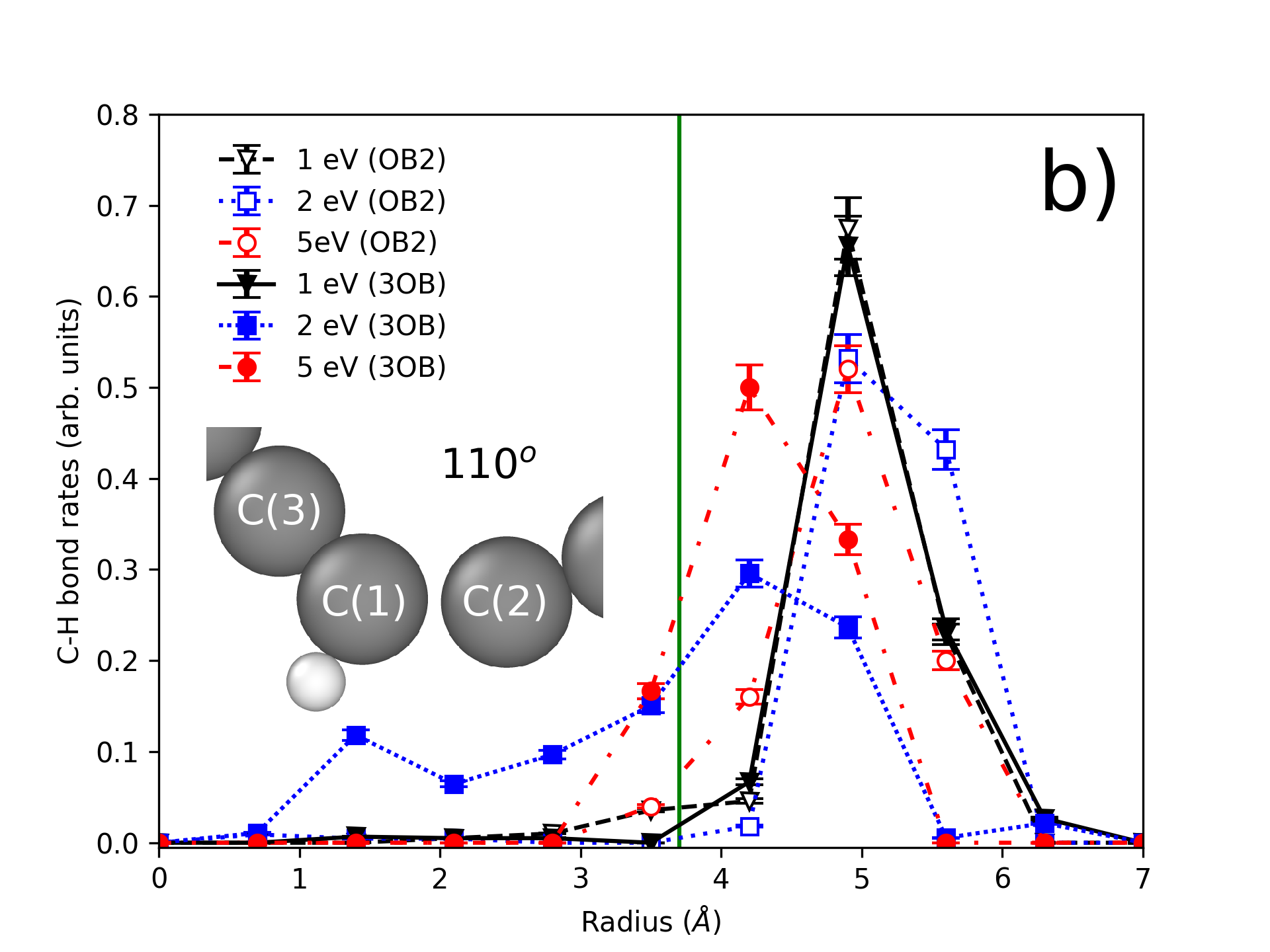}
   \caption{(Color on-line). a) Distribution of the final positions 
   of H atoms bounded to the carbon ring at 2 eV for OB2 and 3OB 
   MD simulations with DFTD3 dispersion corrections. 
   We show the most common final geometry shape of the C$_{18}$
   molecule when a H atom is bound to it. 
   b) The average C-H bond rates as a function of their spatial
   position at 1, 2, and 5 eV for OB2 and 3OB MD simulations. 
   The vertical dotted line at 3.75 \AA{} depicts the radius of 
   the C$_{18}$ molecule. 
   The atomic geometry of the C-H bonding is included where the 
   angle between the C(1)-C(2)-C(3) atoms is $\theta_t= 110^o$ 
   (Carbon atoms are represented by grey spheres and the hydrogen
   atom is shown by a white sphere).}
   \label{fig:fig6}
\end{figure}

In Fig. \ref{fig:fig6} a), we show the position distribution of 
all the events of the H impacts that end up bounded to the carbon
ring at initial impact energy of 2 eV for the OB2 and 3OB MD with the
inclusion of DFTD3 dispersion corrections. 
We find that the most dominant final geometry of the C$_{18}$H
molecule is where one half of the original carbon ring is deformed
by the presence of the hydrogen bound to the C$_{18}$ ring, while 
the second half of the ring preserves its original shape. 
This geometry is observed in all cases where C$_{18}$H is formed 
regardless of the impact energy and SK parameters used in the MD
simulations. 
This confirms that DFTB models properly the $sp$-hybridization of 
the valence electrons, which evolves from  $sp$ to  $sp^2$
hybridization in the process of formation of C$_{18}$H. 
However, the distribution of H atoms bound to the carbon ring 
depends on the SK parameters used in the MD simulations. 
The H atom is more likely to be bound to the C ring outside and
directly to a C atom, as follows from the PECs for the atomic case. 
On the other hand, the 3OB MD simulations show the formation of a 
C$_{18}$H molecule with different position of the H atom:  
The H atom is sometimes bound to the C ring at the inside region,
though still more likely to be bound at the external region. 
In Fig. \ref{fig:fig6} b), we report the C-H bond probabilities 
a function of the distance $R = \sqrt{x^2+y^2}$, where $x$ and $y$ 
are the final position of the H atoms bound to the carbon ring. 
As shown in Fig. 6a) for the 2 eV, OB2 MD simulations (hollow 
symbols) model the formation of the C$_{18}$H molecule with a H 
atom bound outside for different impact energies. 
However, the 3OB MD simulations (solid symbols) agree well with 
those for OB2 at 1 eV. 
The H atom can also be bound to the C ring at the inside region 
for an impact energy of 5 eV according to the 3OB SK parameters. 
At 2 eV, the H atom can be found in a distance range of 1-5 Å. 
We notice that C atoms, in the C$_{18}$H molecule, form an angle
of $110^o$ between the three carbon atoms in the vicinity of the
H atom, as shown in the inset of the figure. 
The C-H bonding modifies half of the geometry of the C$_{18}$ molecule
due to $sp$ hybridization. 
The visualization of the formation of the C$_{18}$H molecule, 
by using 3OB SK parameters with DFTD3 corrections, is provided 
in the supplementary material. 

\subsection{Molecular hydrogen irradiation}

 In order to study the formation of a C$_{18}$H molecule by 
 molecular hydrogen irradiation, we define different events that
 are observed in our MD simulations: The first event is labeled 
 as \textbf{Dyn. 1}, where the H$_2$ molecules are adsorbed by 
 the carbon ring with the hydrogen atoms being bound to one or
 two neighboring C atoms.  
 The formation of a C$_{18}$H$_2$ molecule is observed. 
 In the second event called \textbf{Dyn. 2} the hydrogen 
 molecules are dissociated and one H atom becomes bound to the 
 carbon ring, while the second H atom is scattered.  
 The formation of C$_{18}$H molecules is found. 
 In the third event named as \textbf{Dyn. 3}, kinetic energy of
 the projectile is high enough to break the bond of the H$_2$
 molecule when irradiating the carbon ring. 
 Here the hydrogen molecule is dissociated and scattered while 
 the carbon ring remains in the initial geometry shape. 
 Finally, the \textbf{Dyn. 4}  case, where the most common event
 observed in our MD simulation is the repulsion of the H$_2$ 
 molecule by the carbon ring without dissociating the H$_2$ molecule.
 This event is observed in the whole impact energy range, regardless
 the choice of the SK parameters and level of theory for the
 dispersion corrections. 
 The probabilities of the different events are computed as: 
 $N_E/N_T$ where $N_E$ is the number of cases of each event and 
 $N_T$ is the total number of H$_2$ molecules. 
 The visualization of the dynamics of these events is included in
 the supplementary material.

\begin{figure}[!b]
   \centering
   \includegraphics[width=0.48\textwidth]{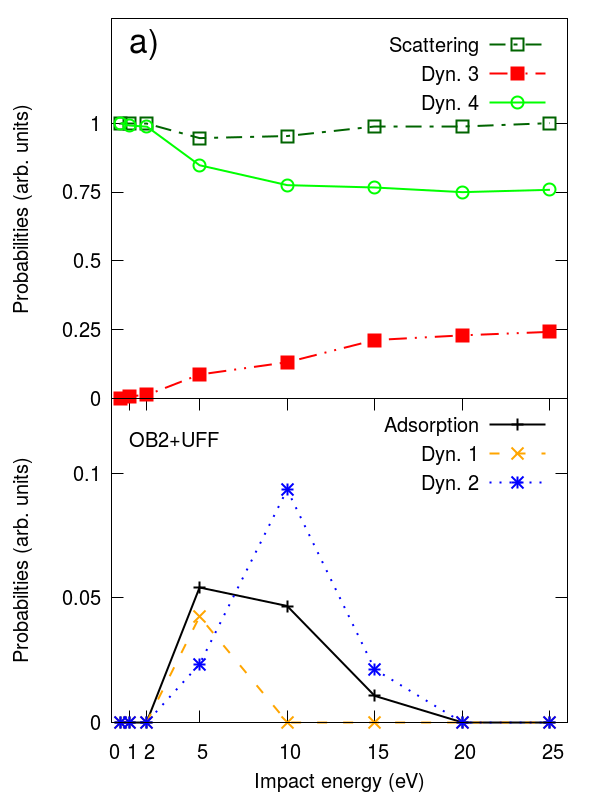}
   \includegraphics[width=0.48\textwidth]{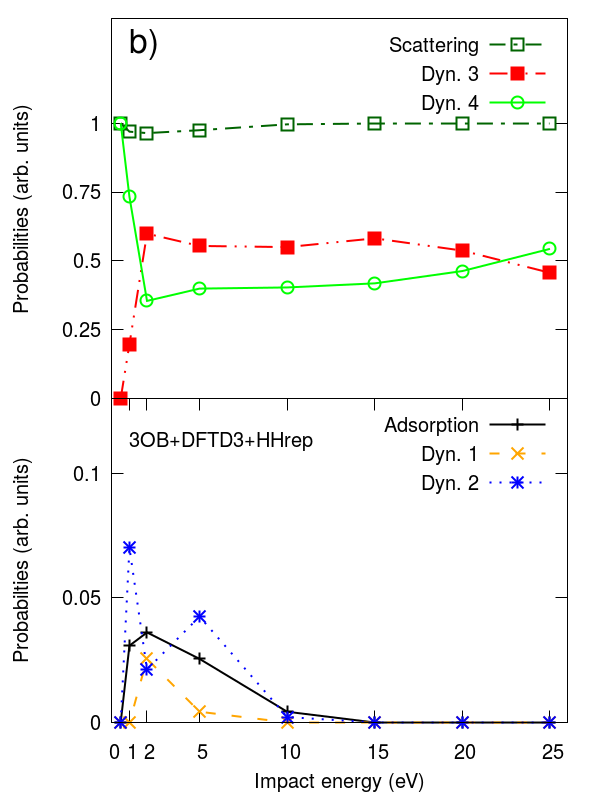}
   \caption{(Color on-line). Probability of adsorption and 
   scattering for H$_2$ molecules as a function of the impact 
   energy.  
   a)  The results for the OB2-UFF approach.
   b) for the 3OB-DFTB3 correction terms. 
   Also shown are the probabilities of different events:
   adsorption
   of a H$_2$ molecule (Dyn. 1); adsorption of a H atom and H
   scattered (Dyn. 2); a H$_2$ molecule is dissociated and
   scattered
   (Dyn. 3); and a H2 molecule is scattered (Dyn. 4).}
   \label{fig:fig7}
\end{figure}

In Fig. \ref{fig:fig7}, we report results for the probability of
adsorption and scattering of H atoms as a function of the impact
energy for the OB2 and 3OB SK parameters in a) and b), respectively.
The probabilities are calculated as follows: $N_{p}/N_{ta}$ with 
$N_p$ the number of scattered or adsorbed H atoms and $N_{ta}$ is 
the total number of H atoms (twice the number of H$_2$ molecules). 
In Fig. \ref{fig:fig7}a), we notice that the H$_2$ molecule is not
bound to the carbon ring for impact energies below 2 eV (lower 
panel), in agreement to the results present by the PECs in 
Fig. \ref{fig:fig4}b). 
H atoms are bounded to the C$_{18}$ molecule in the impact energy
range of 5 to 15 eV, where the H$_2$ impact has enough kinetic energy
to break the 4.5 eV H-H bond. 
For impact energies higher than 2 eV, the scattered H$_2$ molecule 
can be dissociated and this event is dependent on the impact energy.
However, one single H atom of the dissociated H$_2$ molecules is
bound to the carbon ring in an impact energy range of 5-15 eV. 
The event Dyn. 1 is only observed at 5 eV, where a H atom is first
bound to one C atom, leaving the second H atoms free to travel 
around the carbon ring until it is bound to a C atoms at a 
symmetrical position respect to the first H atom. 
Fig. \ref{fig:fig7} b) shows the results obtained by performing 
MD simulations with 3OB SK parameters with DFTD3 dispersion
corrections and considering hydrogen bonding corrections. 
The results for scattered H atoms and molecules differ drastically
with respect to those presented by OB2 SK parameters with UFF
dispersion corrections. 
The probability of scattering reaches a minimum at 2 eV, which is
also an important energy range for the atomic case. 
Probabilities for scattered H atoms do not significantly changes
for impact energies higher than 2 eV. 
The probabilities for the \textbf{Dyn. 3} and \textbf{Dyn. 4} events are similar, 
where the dissociation of a H molecule and the 
reflection/transmission of a H molecule is possible at same 
energies. 
However, the probability of a dissociated and scattered H molecule
is bigger in an impact energy range of 2 to 20 eV. On the other 
hand, the formation of a C$_{18}$H and C$_{18}$H$_2$ molecules is
possible in an impact energy range of 1-10 eV. 
The OB3+DFTD3 results present notable differences when compared 
to those for OB2+UFF calculations. 
The formation of a C$_{18}$H$_2$ molecule is observed at 5 eV for
the latter and at 2 eV for the former. 
This is an effect of the inclusion of HH repulsion contribution 
in the DFTD3 dispersion corrections. 
Here, the molecule is formed when the H$_{2}$ molecule is dissociated
with a H atom bound to a C atom and the second H atoms is bound 
to the first nearest neighbor C atom. 
A C$_{18}$H can be formed at impact energies between 1 and 5 eV, 
as reported by the 3OB+DFTD3 results.

\begin{figure}[!b]
   \centering
   \includegraphics[width=0.48\textwidth]{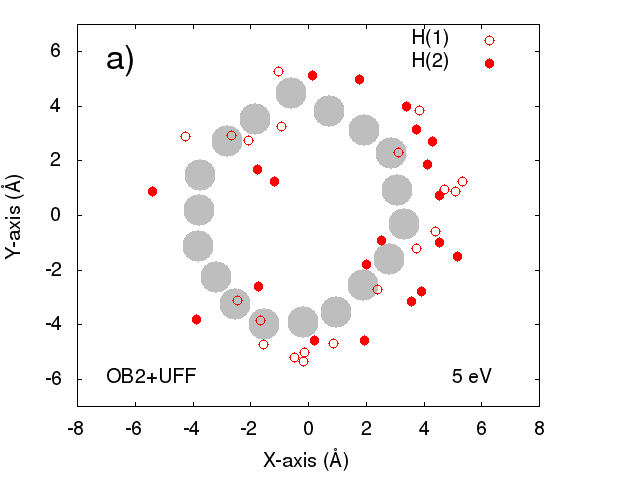}
   \includegraphics[width=0.48\textwidth]{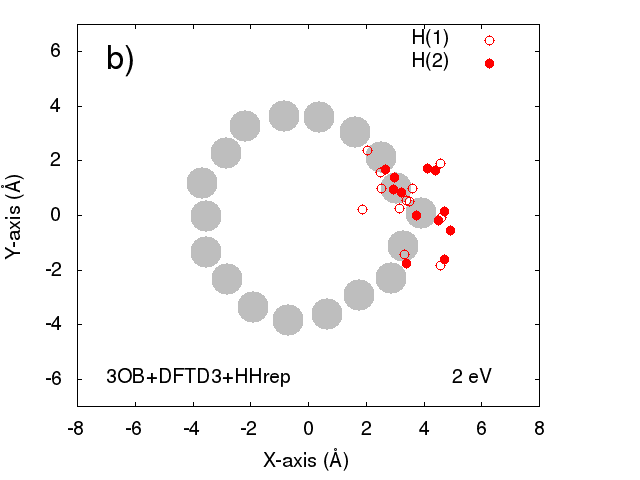}
   \caption{(Color on-line). Distribution of the final position of H 
   atoms that form a C$_{18}$H$_2$ molecule as obtained by
   OB2+UFF in a) and by 3OB+DFTD3 in b) after irradiation dynamic by H$_2$. 
   We include the most common final shape of the carbon ring. 
   For the formation of C$_{18}$H$_2$ molecules, we differentiate
   the hydrogen atoms bound to the C$_{18}$ molecule according to 
   the {\bf Dyn. 1} event, where the H$_2$ is dissociated and 
   the H atoms are bound to the carbon ring.}
   \label{fig:fig8}
\end{figure}

In Fig. \ref{fig:fig8}, we present all cases for the formation 
of a C$_{18}$H$_2$ molecule. 
In Fig \ref{fig:fig8} a), we show the results for the OB2+UFF and 
in \ref{fig:fig8} b) for 3OB+DFTD3, according to the {\bf Dyn. 1} and
{\bf Dyn. 2} events. 
Attention needs to be paid to the H atoms at the inside region 
of the carbon ring as the hydrogen atom is not located in the C 
ring. 
We also include the most common shape of the carbon ring at the 
final time of the MD simulations for both cases. 
Notice that the carbon ring is completely deformed by the binding
of two H atoms for the OB2+UFF case. 
Interestingly, the H atoms are symmetrically bound to two 
different carbon atoms, one at one side of the ring and the other
at opposite side. This produces a semi-elliptic shape of the 
C$_{18}$H$_2$ ring. 
Results for the 3OB+DFTD3 case present a formation of the 
C$_{18}$H$_2$ molecule in a systematic way. 
The H atoms are bounded to the first nearest C atom and the 
final shape of the carbon ring is barely affected. 
Only one half of the ring is deformed when the H atoms are bounded
to the C atoms. 
Besides, the H atoms are mainly bounded to the C ring at the 
outside region, in good agreement with the atomic case. 
Of main importance is that we haven’t observed fragmentation of
the C$_{18}$ molecule by irradiation of hydrogen. 
This shows that the C$_{18}$ molecule ring is stable and thus a 
good candidate for hydrogen absorption or detection. 
The formation of C$_{18}$H$_{2}$ molecule suggests the use of 
the 3OB+DFTD3 parameters for a study of the formation of a 
C$_{18}$H$_n$ molecule, with $n > 2$ in a cumulative process, 
which is work in progress and we will report its outcome 
elsewhere.

%%%%%%%%%%%%%%%%%%%%%%%%%%%%%%%%%%%%%%%%%%%%%%%%%%%%%%%%%%%%%%%%%%%%%%%%%%%%%%%%%%%%%%%%%
\section{Concluding remarks}
\label{sec:Concl.}
In this work, we performed numerical simulations with a
molecular-dynamics approach based on the DFTB method to 
simulate irradiation by atomic and molecular hydrogen of 
a C$_{18}$ molecule (cyclo[18]carbon). 
We consider different sets of SK parameters to model the 
inter atomic interaction between the C and H atoms, as 
well as different approaches to include van der Waal 
interactions in our atomistic simulations by dftb$+$. 
We have used Slater-Kirwood polarizable atomic model, 
Lennard-Jones potential with Universal Force Field 
parameters, and the 3-body dispersion energy through a 
damped pairwise London-type factor correction in the 
DFTB approach. 
Potential energy curves (PECs) for the interaction of 
C$_{18}$ with atomic and molecular hydrogen at different 
adsorbate sites are computed to test the suitability of 
the DFTB parameters to properly model the physical and 
chemical physio-absorption process during the hydrogen 
irradiation on the carbon ring. 
These PECs also serve as a guide to understand the MD 
simulation results. 
For atomic irradiation, we found that the C$_{18}$H 
molecule is more likely to be formed at impact energies 
below 10 eV, regardless the choice of the DFTB parameters 
and dispersion corrections. 
However the collision dynamics differ between the DFTB 
parameters at very low impact energies, where the H atom 
is mainly bound to the carbon at the outer region of the 
molecule for the case of the OB2 parameters. 
In contrast, the recently developed 3OB parameters model 
a C$_{18}$H molecule with a H atom bound at the inner region 
for impact energies higher than 1 eV. 
For the molecular case, we found that the formation of a 
C$_{18}$H$_{2}$ molecule is feasible in an impact energy 
range between 2-15 eV. 
However differences are observed in the MD simulation 
results for OB2 and 3OB parameters. 
The formation of this molecule is only observed around 5 eV 
for the OB2 case and around 2 eV for the 3OB one. 
This disagreement is due to the inclusion of a DFT 3-body 
level of theory in the dispersion corrections and 
hydrogen-hydrogen repulsion corrections for the 3OB MD 
simulations with the DFTB3 approach. 
In conclusion, we find that the hydrogenation of C$_{18}$ 
by atomic and molecular hydrogen irradiation is highly 
probably at low irradiation energies in a single collision 
event. 
Work is in progress to study the feasibility of multiple 
hydrogen binding, either sequentially or in a cumulative 
process. 

%%%%%%%%%%%%%%%%%%%%%%%%%%%%%%%%%%%%%%%%%%%%%%%%%%%%%%%%%%%%%%%%%%%%
\section*{Acknowledgments}
%\ack
FJDG acknowledges support from the A. von Humboldt foundation for research fellowship. 
RCT acknowledges support from DGAPA-\-UNAM PAPIIT-\-IN-\-111-\-820 and LANCAD-\-UNAM-\-DGTIC-\-228. 
CMF thanks CONACyT for the postdoctoral fellowship through the project FC-2016/2412.
We also thank to Prof. Marcus Elstner from the Karlsruhe Institute of Technology for 
fruitful disussion about the DFTB method. 
Results in this paper were obtained using the Seawulf institutional cluster at the Institute 
for Advanced Computational Science in Stony Brook University and the 
Max-Planck Computing and Data Facility.

%%%%%%%%%%%%%%%%%%%%%%%%%%%%%%%%%%%%%%%%%%%%%%
%\appendix
%\section{Supplementary material}
%\input{sections/supp_mat}
%%%%%%%%%%%%%%%%%%%%%%%%%%%%%%%%%%%%%%%%%%%%%%%%%%%%%%%%%%%%%%%%%%%%%%%
%\References
%\section*{References}
\bibliographystyle{iopart-num}
\bibliography{bibliography}	

\end{document}